\documentstyle[aps,prl,twocolumn,psfig]{revtex}
\begin{document}


\newcommand{\tit}
{Finite size effects and the mixed quark-hadron phase in neutron stars}
\newcommand{\authb} {Norman K. Glendenning}
\newcommand{\autha} {Michael B. Christiansen}
\newcommand{\authc} {F. Weber}
\newcommand{\lbl}{\begin{flushright} LBL-40284\\[7ex] \end{flushright}}
\newcommand{\dateofdoc}{\today}
\newcommand{\doe}
{This work was supported by the
Director, Office of Energy Research,
Office of High Energy
and Nuclear Physics,
Division of Nuclear Physics,
of the U.S. Department of Energy under Contract
DE-AC03-76SF00098.}

\newcommand{\ect}{A part of this work was done at the ECT*,
Villa Tambosi, Trento, Italy.}

\begin{titlepage}
\lbl
\begin{center}
\begin{Large}
\renewcommand{\thefootnote}{\fnsymbol{footnote}}
\setcounter{footnote}{1}
\tit {\footnote{\doe}}\\[5ex]
\end{Large}

\renewcommand{\thefootnote}{\fnsymbol{footnote}}
\setcounter{footnote}{2}
\begin{large}
\autha~~~ \authb \\[3ex]
\end{large}
\dateofdoc \\[3ex]
\end{center}


\begin{figure*}[tbh]
\vspace{-.5in}
\begin{center}
\leavevmode
\hspace{-.8in}
\psfig{figure=ps.fig2,width=4.in,height=4.in}
\end{center}
\end{figure*}

\begin{center}
{\Large Institute of Physics and Astronomy\\
University of Aarhus, Denmark\\and\\
Nuclear Science Division and Institute for Nuclear
\& Particle Astrophysics\\ Lawrence Berkeley National Laboratory\\
Berkeley, California}
\end{center}
\end{titlepage}

\clearpage
\draft
\title{Finite size effects and the mixed quark-hadron phase
in neutron stars} 
\author{Michael B. Christiansen}
\address{Institute of Physics and Astronomy,
University of Aarhus, 
DK-8000 \AA rhus C, Denmark} 
\author{Norman K. Glendenning}
\address{Nuclear Science Division and
Institute for Nuclear \& Particle Astrophysics,  
 Lawrence Berkeley Laboratory, MS: 70A-3307, Berkeley, California 94720}
\date{\today}
\maketitle

\begin{abstract}
We demonstrate that the form and location (not the size or spacing) of the 
energetically preferred geometrical 
structure of the 
crystalline quark-hadron mixed phase in a neutron star is very sensitive to
finite size terms beyond the surface term. We consider two independent 
approaches of including further finite size terms, namely the multiple 
reflection expansion of the bag model
and an effective surface tension description. 
Thus care should be taken in any model requiring detailed knowledge of these 
crystalline structures. 

\end{abstract}


\section{INTRODUCTION}
The possible existence of quark matter in the dense cores of neutron stars
was proposed in pioneering work two decades ago 
\cite{baym76,keister76,chapline76} and has been
further studied by a number of authors 
\cite{kisslinger78,freedman78,joss78,serot87,kapusta90,kapusta91}.
The theory has recently been reexamined
and new insights into the nature of the phase transition were gained which 
drastically affect the structure of neutron stars \cite{norm92}. 
In  previous work
the mixed or coexistence phase was strictly excluded from the
star and a large density discontinuity occured at the radial point 
corresponding to the (constant) pressure of the mixed phase. It is now
understood that   the common pressure and all properties of the
two phases in equilibrium vary as their proportion. Consequently
the mixed phase can occupy a large radial extent of some kilometers
dimension between the two pure phases. Moreover, the mixed phase region
is expected to achieve an energy 
minimum by formation of a crystalline lattice of the
rarer phase immersed in the dominant one. The dimensions and 
geometrical forms of the lattice
will vary with the proportion of the phases and hence with
depth in the star \cite{norm92}.
The discontinuity in density of the early work is replaced by a region
made of an intricate pattern of crystalline forms.

The assumption underlying the previous picture
is that the confining
transition from quark matter to hadrons is a first order transition with
effectively one
conserved charge, namely the baryon number. This situation arose
in some papers through treatment of the star
as being made solely of  neutrons (which is
beta unstable and therefore an excited state of the
star) and in others by the imposition
of {\sl local} charge neutrality on a beta
stable system which effectively also reduced the
phase transition
to  one with a single conserved charge \cite{norm92}.

In either case therefore, the transition was made to resemble that of the
liquid-vapor transition 
in water, a constant pressure-temperature one.
However first order phase transitions in substances of two or more
independent components such as neutron stars in beta equilibrium (for which
baryon number and electric charge number are the conserved quantities)
behave quite differently \cite{norm92}. This follows
since the conservation laws are {\sl not} local ones but are global.
In fact imposition of local neutrality violates Gibbs conditions for
phase equilibrium.
When the two phases are in equilibrium with each other they can 
rearrange the concentration of the conserved charges to optimize the
total energy while conserving the charges overall. In general, energy
will be minimized by a different concentration of
conserved charges at each proportion of the
phases; hence all properties of the phases also vary with proportion.

In nuclear systems it is the isospin restoring interaction 
(responsible for the valley of beta stability in nuclei)
that exploits the degree of freedom available by the possibility of
rearranging concentrations while conserving charges. The consequence is that
regions of confined and deconfined matter will have opposite
charge densities and a structured Coulomb lattice minimizes the energy.
Elsewhere it was found  that
the unstructured phase would be preferred
if the surface tension where too large \cite{heipet}. We have  misgivings
about this  conclusion however, which we outline next. 

In Ref.\ \cite{heipet}
the surface tension was treated as a free parameter, which it
is not. It was therefore
possible to find a value that was large enough to put the
structured phase at a higher energy than the unstructured one.
However, opening a degree of freedom has the effect of lowering energy,
or leaving it unchanged, not raising it. 
Were it possible do an exact calculation of
phase equilibrium including the possibility of spatial structure,
without introducing the bulk approximation corrected for Coulomb energy,
surface energy,
curvature, etc, the above physical principle would be
obeyed. It is therefore incorrect in the approximation scheme
(bulk+surface+coulomb+curvature)
to arbitrarily choose
a value of surface tension that places the structured
phase above the unstructured one.

The form, size, spacing and location in the star of the crystalline mixed 
phase was first calculated by Glendenning and Pei \cite{pei}, where the sum 
of Coulomb and surface energies was minimized as a function of the proportion
of the two phases in equilibrium. Here we extend their calculation to include 
further finite size effects, i.e.\ curvature energy and even higher order terms,
for the three simple geometries, spheres, rods and slabs. We do this in two 
completely independent ways, and show that the inclusion of additional finite
size effects has a significantly larger influence on the 
preferred geometrical form, than on the actual size of the form.

In Sec.\ II we give a general description of the mixed phase. 
Sec.\ III and IV are devoted to a description of the two different approaches
to include further finite size corrections, and 
Sec.\ V contains a general discussion and a summary of 
our results.

\section{THE MIXED PHASE}

The equation of state of the confined hadronic phase is calculated as
 in Ref.\ \cite{glen85:b,glen91:c}.
 The coupling constants of the theory are determined by the
 bulk nuclear properties, binding energy,
 $B/A=-16.3$ MeV, saturation density, $\rho_0=0.153$ fm$^{-3}$,
 symmetry energy coefficient, $a_{{\rm sym}}=32.5$ MeV,
 compression modulus $K=240$ MeV
 and effective nucleon mass at saturation, $m^\star/m=0.78$.
 The ratio of hyperon coupling to mesons as compared to nucleon couplings
 is chosen in accord with Ref.\ \cite{glen91:c} to be $x_\sigma=0.6,~
 x_\omega=0.658$.

The bulk properties of the quark phase is described in the bag model at zero 
temperature with the strong coupling constant, $\alpha _s=0$. The relevant 
expressions for the pressure, energy density, baryon number and charge density
can be found in Ref.\ \cite{norm92}.
Throughout, we use a bag constant, $B^{1/4}=180$ MeV. We take for the quark
masses $m_u=5$ MeV, $m_d=10$ MeV and $m_s=150$ MeV. No heavier quarks are 
present. While the importance on the bulk properties from using non-zero $u$ 
and $d$ quark masses is negligible, the surface tension is increased about 
15 percent in the so-called thin wall limit (only massive quarks contribute 
to the surface tension). 

The relation between the pressure and energy density, and the neutron star's
Schwarzschild radius is found from solving the Oppenheimer-Volkoff equations.
Moving inwards through the star, the mixed phase becomes energetically favorable
when the energy density has increased to a value where the bulk pressures of 
the two phases have become equal. 
In our approximation (volume+surface+Coulomb+....)
the formation of geometrical structures in the mixed phase only occurs if the
sum of finite size and Coulomb energy densities is less than about 10 
MeV/fm$^3$. 
In the next section, where finite size corrections are determined from the
bag model this does not pose a problem as the sum of energy densities is at 
most 7 MeV/fm$^3$. In Sec.\ IV, where the surface energy cannot be explicitly 
calculated from the model, 
we ensure that the sum does not exceed 10 MeV/fm$^3$ by appropriate 
adjustment of the surface tension.
The geometrical structure of the mixed phase occurs to a good approximation
against the background of the bulk structure, except in the outermost regions
of mixed phase,
where the surface pressure from the geometrical structures are comparable with
the bulk pressures. 
The Debye screening lengths were estimated in Refs.\ \cite{heipet,screen} to 
be about 5 fm. This is approximately the typical radius of the geometrical 
structures and thus screening is also only a small effect.
We will neglect these minor complications. 
If the neutron star is massive enough it will have a pure core of quark matter.
We focus on a neutron star at the mass limit, which with our choice for the 
equation of state is at $1.454 M_{\odot}$ and a radius of 10.32 km. 

The mixed phase is subdivided into Wigner-Seitz cells, that have total charge 
zero. Adjacent cells therefore do not interact. Each cell has the same form 
as the one structure of rare phase it contains. The surrounding region of the
other phase is treated as a bulk system. 
For the simple discrete geometric forms described in \cite{raven,pei}, drops,
rods and slabs, each characterized by the dimensionality $d$ equal to 3, 2, 
and 1, respectively,
the Coulomb energy per volume for the three structures can be written as 
\begin{equation}
 E_C = 2\pi \left [ \rho _h(\chi) -\rho _q(\chi) \right ]^2 r^2 x f_d(x)
     = C(x) r^2,
\end{equation}
where $\rho _h$ and $\rho _q$ are the charge densities of the hadronic and 
quark matter at volume proportion $\chi =V_q/V$ of quark matter. The ratio of
structure volume to cell volume is denoted by $x=(r/\cal R$)$^d$. When 
quark matter 
is the rare phase, $x=\chi$, and otherwise, $x=1-\chi$. In the case of drops 
or rods, $r$ is their radius and $\cal R$ the half distance between centers, 
whereas for slabs, $r$ is the half thickness.
The function $f_d(x)$ is in all three cases given by 
\begin{equation}
 f_d(x)=\frac{1}{d+2} \left [ \frac{1}{d-2} \left ( 2-d x^{1-2/d} \right ) 
       + x \right ].
\end{equation} 

The preferred geometrical form is found by minimizing the sum of finite size 
and Coulomb energies. We take two different approaches to get expressions for 
the finite size corrections beyond the surface term.

\section{FINITE SIZE CORRECTIONS IN THE BAG MODEL}

From the multiple reflection expansion method \cite{balian} used with MIT
bag boundary conditions \cite{johnson}, the finite size corrections 
to the energy can 
be explicitly calculated. This has been done up to curvature corrections for
massless quarks, but the ansatz provided by Madsen \cite{madsen} for the 
curvature 
term for massive quarks has shown to reproduce exact mode-filling calculations
for quarks at zero temperature.
The use of MIT bag boundary conditions imply that the thickness of the boundary 
wall is zero, the so-called thin wall approximation. In this approximation 
the two phases in equilibrium are treated independently, i.e.\ any finite size 
energy contributions from the hadronic phase just add to the quark matter 
contributions.
The surface energy per volume for the quark matter is
\begin{equation}
 E_S= \frac{d x \sigma}{r}=\frac{S(x)}{r},
\end{equation}
and the curvature energy per volume
\begin{equation}
 E_{\Gamma}= \pm \frac{d(d-1)x\gamma }{r^2} =\pm \frac{\Gamma (x)}{r^2},
\end{equation}
where ``+'' is used when $x=\chi$, and ``$-$'' when $x=1-\chi$. The curvature 
term vanishes of course for slabs.  
Expressions for the surface tension, $\sigma$, and curvature coefficient, 
$\gamma$, can be found in Ref.\ \cite{madsen}. $\sigma$ decreases from about 37 
to 25 MeV/$\hbox{fm}^2$, and $\gamma$ decreases from 25 to 12 MeV/fm as the 
chemical potentials drop out through the star. The nuclear liquid drop model
give a surface tension $\sigma _{had} \simeq 1$ MeV/$\hbox{fm}^2$ for ordinary
nuclear matter in vacuum that, by comparison with the quark contribution, 
can be ignored in the thin wall approximation \cite{myers}.

When the quark phase is the rare one, the sum of finite size and 
Coulomb energy densities diverges to plus infinity as a function of $r$ in both
the $r\rightarrow 0$ and $r\rightarrow \infty$ limits, and thereby ensuring 
the existence of a local energy minimum. But the sign change in the curvature
energy for a bulk quark phase has devastating consequences, as the sum of 
energy densities diverges to minus infinity for vanishing $r$. 
First we cannot be sure 
there is a local minimum, and even worse the energy can be lowered by creating
more and more smaller and smaller hadronic bubbles in the quark matter.
The sign change in the curvature energy is not a pathological defect of the 
model. The curvature energy is proportional to the mean curvature integrated
over the surface of the geometry, and is therefore a signed quantity unlike 
the surface area. This has explicitly been confirmed by Mardor and Svetitsky
\cite{mardor91} in the high temperature, zero chemical potential limit.
We take this as a warning of how important it is to ensure that enough terms
in the expansion are included. 

The next term in a finite size expansion of the energy, at least for 
massless quarks, 
is of the form $Z_o/r$ for spheres, and is identified as the Casimir or 
zero-point energy. Unfortunately, $Z_o$ has
not been evaluated in the multiple
reflection expansion of the bag model. However, similar procedures,  
give $Z_o \sim 1$ \cite{francia,plunien}. 
Bag model fits to the hadronic spectrum are considerably improved if a 
zero-point term of $-1.84/r$ is added to the energy \cite{degrand75}. About half
of it can be explained in terms of center of mass corrections from localization 
of the hadron \cite{com}.
We therefore choose to treat $Z_o$ as 
a phenomenological parameter of order one for all three geometries. 
For the bulk quark phase, where $E_{\Gamma}<0$, we will insist upon $Z_o$ 
being positive; whereas in the bulk hadronic case, where $E_{\Gamma}>0$, we 
will allow both signs for $Z_o$. Even if $Z_o$ is negative, the sum of finite 
size and Coulomb energies will not be negative except for radii so small that 
we do not believe in the expansion anyway.
This always ensures the existence of a local energy minimum.

For cylinders and parallel plates the zero-point terms are
\cite{milton,plunien}
$E_{cyl}=Z_ol/r^2$ and $E_{plate}=Z_ol^2/a^3$,
respectively. $l$ is the length of the structure, $r$ the cylinder radius, and
$a$ the distance between the plates, in our notation $2r$. The zero-point 
energies per volume are then 
\begin{equation}
 E_Z= \frac{Z(x)}{r^4} = \frac{Z_ox}{r^4} \times \left\{ \begin{array}{ll} 
       \frac{1}{16}, & d=1, \\[.5ex]
       \frac{1}{\pi}, & d=2, \\[.5ex]
       \frac{3}{4\pi}, & d=3. 
                                \end{array} \right. 
\end{equation}
The sum of finite size and Coulomb energies can be written as 
\begin{equation}
 E= C(x)r^2+\frac{S(x)}{r} \pm \frac{\Gamma(x)}{r^2} + \frac{Z(x)}{r^4}.
\label{sum}
\end{equation}
The minimization procedure is carried out numerically for the three geometries
in consideration, under the constraint that $2E_C=E_S + 2E_\Gamma + 4E_Z$.
\begin{figure}[tbh]
\vspace{-.2in}
\begin{center}
\leavevmode
\psfig{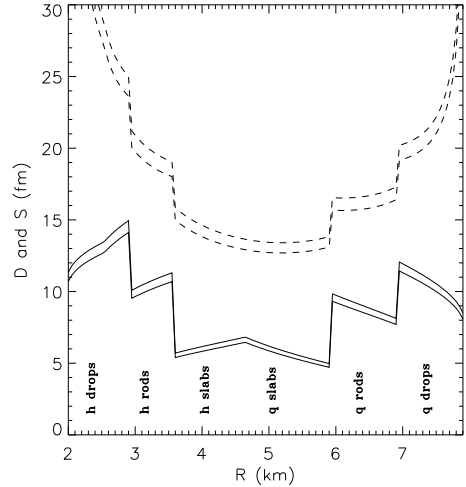}
\parbox[t]{5.5in} { \caption { \label{fig1} Mixed phase
region in a neutron star showing the diameter $D$ (solid
curves) and spacing $S$ (dashed curves) as a function of the radial
Schwarzschild coordinate. The curves on top are for massive $u$ and $d$ quarks,
in the bottom ones they are assumed massless. ``$h$ drops'' denotes hadronic
drops immersed in bulk quark matter, and so forth.
}}
\end{center}
\end{figure}

We plot, as a function of the radial Schwarzschild coordinate, $R$,  
the diameters of the different geometric structures 
present in the mixed phase region of the 
neutron star, and the spacing between adjacent structures immersed in the bulk 
phase.  
The core of the neutron star up to $R=2.0$ km consist of pure quark matter, 
while beyond $R=7.9$ km the neutron star is composed of ordinary hadronic 
matter. 
In Fig.\ \ref{fig1} we have neglected the curvature and zero-point energies 
and show the increase in structure size of about 5\%, when the $u$ and $d$ 
quarks are not assumed massless.
All six geometric forms are present in this plot. A new form becomes 
energetically favorable at each non-differential point on the diameter 
curve. The discontinuities are only present because of the discrete 
geometries we treat. Their location is independent of the surface tension, 
only the radii of the geometries are affected. The radii lie between 2.5 and 
7.5 fm, and therefore screening effects are only of minor importance even for 
the largest structures.

\begin{figure}[tbh]
\vspace{-.2in}
\begin{center}
\leavevmode
\psfig{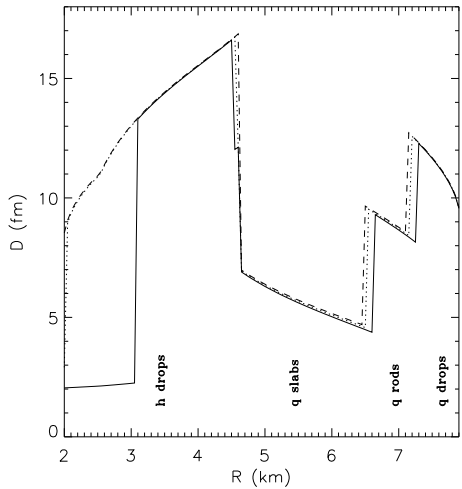}
\parbox[t]{5.5in} { \caption { \label{fig2}Similar to Fig.\ \protect\ref{fig1}
but
without the spacing curves. The
solid, dotted and dashed curves are for $Z_o$ equal to 1, 1.5 and 2,
respectively. $Z_o$ is assumed positive in both bulk phases.
}}
\end{center}
\end{figure}

Fig.\ \ref{fig2} is a plot similar to Fig.\ \ref{fig1} without the spacing 
curves. Included is now the curvature energy and three zero-point energies,
assumed positive in both bulk phases.
In comparison with Fig.\ \ref{fig1} two striking new features emerge. First,
the solid diameter curve, that represent the smallest zero-point energy, 
$Z_o=1$, abruptly drops at $R\simeq 3$ km, but without changing the preferred 
geometry, hadronic drops. 
The reason is that a second minimum in the sum of finite size and Coulomb 
energies, Eq.\ (\ref{sum}), show up at around $r=1$ fm, if the zero-point term 
is small enough.
For increasing $Z_o$ the second  minimum will eventually not be the global 
one, this happens when $Z_o>1.5$, and will finally even disappear. 
Even though the sum of energies is positive for $Z_o=1$ as it should be, we
neither expect nor believe the model to be valid for such small radii. So we
must take $Z_o>1.5$. 
Second, no hadronic slabs are present. This is no big 
surprise, since the curvature energy is zero for the slabs, while negative for
hadronic drops and rods. Hadronic rods are only preferred in a very narrow 
range when $Z_o$ is unacceptably small. Thus the inclusion of further 
finite size 
corrections has large implications for the location, and even presence of the 
geometric structures in both bulk phases;
whereas the sizes are unnoticeably affected, except for small $Z_o$, where we 
do not believe in the model.  

\begin{figure}[tbh]
\vspace{-.2in}
\begin{center}
\leavevmode
\psfig{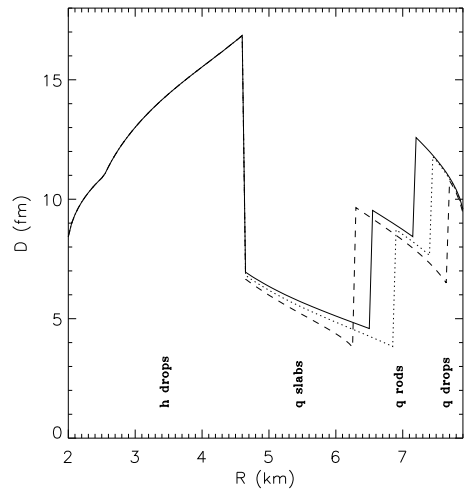}
\parbox[t]{5.5in} { \caption { \label{fig3}Similar to Fig.\ \protect\ref{fig2}.
The solid, dotted and dashed curves
are for $Z_o$ equal to 1.6, 0 and -1.6 in the bulk hadronic phase. $Z_o=1.6$
in the bulk quark phase.
}}
\end{center}
\end{figure}

In Fig.\ \ref{fig3}, which is similar to Fig.\ \ref{fig2}, we have fixed the
zero-point energy for the bulk quark phase and show how the location and size
of the preferred geometry changes for three different values of $Z_o$ for the 
bulk hadronic phase \cite{remark1}. 
For the bulk quark phase the three curves are of course identical. Again
we see that the structure size is relatively less affected by the change in 
$Z_o$ than the location of the geometry. The radial extent of the quark drops
decrease with decreasing $Z_o$. The division between quark slabs and rods does
not exhibit the same simple behaviour.

\begin{figure}[tbh]
\vspace{-.2in}
\begin{center}
\leavevmode
\psfig{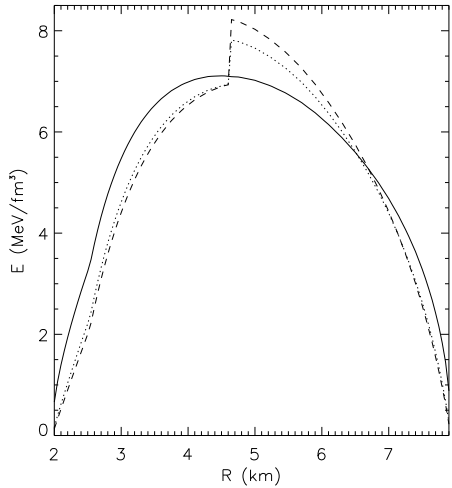}
\parbox[t]{5.5in} { \caption { \label{fig4}The sum of finite size and Coulomb
energies as a function of
Schwarzschild coordinate for $Z_o=1.6$ in both bulk phases. The solid,
dotted, and dashed curves represent slabs, rods, and drops, respectively.
}}
\end{center}
\end{figure}

We have assumed that $Z_o$ for all three geometries are identical, at least in
each of the two phases. We have no reason to expect this should be correct, 
but the assumption reduces the possible six different $Z_o$'s to two,
still enough to show the sensitivity of the location of a geometrical 
structure to the parameter choice.
If the small contribution from the hadronic phase is added to the surface 
energy, only a 3-4\% increase in the surface tension, and the outcome plotted 
in Fig.\ \ref{fig3}, the only noticeably 
difference is that hadronic rods are now preferred in a narrow 
range between hadronic drops and quark slabs, the radii are not visibly 
affected. Furthermore, the smallest 
acceptable value for $Z_o$ is reduced to 1.3.
Fig.\ \ref{fig4} shows the reason for this sensitivity. In this figure 
the sum of finite size and Coulomb 
energies have been plotted as a function of the radial coordinate for each 
of the three geometries. The discontinuity at $R=4.6$ km is solely due to 
the sign change in the curvature energy. The zero-point term is positive in
both bulk phases. Notice especially how close the rod and drop curves lie in 
most regions. It should be obvious that even small changes in parameters 
could displace the locations of the preferred geometries. 

\section{THE MIXED PHASE IN AN EFFECTIVE SURFACE TENSION DESCRIPTION}

In connection with the cosmological phase transition, Kajantie {\it et al.}
\cite{potvin93} have derived an expression for the effective surface 
tension for a diffuse interface between quark and
hadronic matter in equilibrium, based on a simple Taylor expansion of the 
surface tension of a planar surface. 
Also in the high temperature regime the curvature term is negative in the 
bag model for hadronic bubbles in bulk quark matter \cite{mardor91}. The 
sign and magnitude has been confirmed by lattice QCD calculations  
\cite{lattcurv}. This inevitably leads to a negative local minimum in the free 
energy even above the phase transition temperature. 
The model of Kajantie {\it et al.} has the attracting feature of reproducing
the negative curvature term 
when the expression for the effective surface tension is expanded, but
still higher order terms prevent the puzzling negative local energy minimum 
from ever forming.

We cannot immediately use the results for the effective surface tension 
presented in that
paper. The reason being that in our case the pressures and the surface tension 
are functions of two variables, namely the baryon chemical potential, $\mu_n$,
and the electron chemical potential, $\mu_e$, instead of being just a function
of temperature. This reflects the presence of two conserved charges in our 
case. 
But because the electron distribution is uniform and $\mu_e\ll \mu_n$, the
effective surface tension can still be written in the form, generalized to 
also include cylindrical and planar geometries (see Appendix A),
\begin{equation}
 \sigma (r) \simeq \frac{\sigma (\infty)}{1 \pm (d-1)\delta/r}, \qquad 
 \frac{(d-1)\delta}{r}\ll 1,
\label{sigma_r}
\end{equation}
where ``+'' is used when $x=1-\chi$ and ``$-$'' when $x=\chi$.
$d$ is the dimensionality, $\sigma(\infty)$ the surface tension of a planar 
surface, and $\delta$ is a parameter of dimension length given as
\begin{equation}
 \delta = \frac{\left ( \partial \sigma / \partial \mu_n 
 \right )_{\mu^c_n,\mu^c_n}}{n_{n,q}-n_{n,h}},
\label{delta}
\end{equation} 
where $n_{n,i}$ is the baryon number density for each of the two phases.
  
We take, as in Ref.\ \cite{pei}, for the surface tension 
\begin{equation}
 \sigma(\chi) = {\rm const} \times
 \left [ \epsilon_q(\chi)-\epsilon_h(\chi) \right ]
 L,
\label{sigma}
\end{equation}
\begin{figure}[tbh]
\vspace{-.2in}
\begin{center}
\leavevmode
\psfig{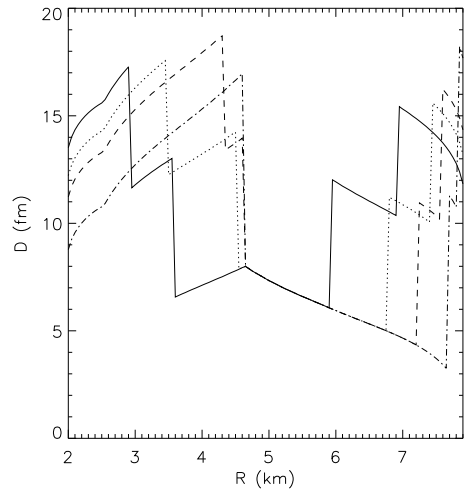}
\parbox[t]{5.5in} { \caption { \label{fig5}Similar to Fig.\ \protect\ref{fig2}.
The curves are for $\delta$ equal to
0 (solid), 0.5 (dotted), 0.9 (dashed), and 2 fm (dash-dotted), respectively.
}}
\end{center}
\end{figure}
where $\epsilon_i$ is the energy density for each of the two phases. $L$ is 
a length scale of the order of the interface thickness, about 1 fm. The 
constant is chosen so that the sum of finite size and Coulomb energy densities 
does not exceed 10 MeV/fm$^3$. With this choice for the surface tension 
($\sigma$ varies between 51 and 66 MeV/fm$^3$),
$\delta$ can actually be calculated, but for our purpose of demonstrating 
the large sensitivity of the location of a geometrical structure to the 
parameter choice, we will treat $\delta$ as a free parameter. Estimates based 
on Eq.\ (\ref{delta}) show that
$\delta(\chi)$ varies between a fraction of one fermi to a few fermi's, but 
in some regions around $\chi\simeq 0.5$ it may actually become negative.

We plot in Fig.\ \ref{fig5} the structure diameter as a function of the 
radial coordinate for four different values of $\delta$, where Eq.\
(\ref{sigma_r}) is valid.
The solid curve where $\delta=0$, i.e.\ no finite size terms beyond the 
surface term, is similar to the solid curves in Fig.\ \ref{fig1}. 
The maximum structure radii are somewhat larger than in the bag model, 
because of the larger surface tension. 
When $\delta$ is increased, the hadronic drops will take up an increasing 
fraction of the bulk quark mixed phase region at the expense of hadronic rods
and slabs, both of which finally even disappears within the range of the 
chosen $\delta$'s.    
For the bulk hadronic matter it is the quark slabs that take up an increasing
fraction at the expense of quark rods and drops. 
If $\delta$ is negative, the general tendencies become inverted, quark drops
and hadronic slabs will eventually dominate the mixed phase region.

Also in this model the structure sizes are relatively less affected, than the 
location and even presence of a geometric structure. This is expected 
because the general trends are of course identical in the two approaches of 
including further finite size corrections, as the leading contribution comes 
from the curvature term, which are similar in the two approaches. 
The curvature term calculated from the bag model is reproduced for 
$\delta \approx 0.6$, but of course additional terms are different in the two
approaches.

\section{CONCLUSION}

We have used two different models to include finite size corrections beyond
the surface term in the sum of finite size and Coulomb energies.
Within the multiple reflection expansion of the MIT bag model the surface 
tension and curvature coefficient can be explicitly calculated. Because of the
negative curvature term, present when hadronic matter is immersed in bulk 
quark matter, we were forced to add an additional term, which was identified 
as a zero-point energy stemming at least partly from Casimir effects.  
The zero-point term was treated as a phenomenological parameter, since terms
beyond the curvature term have
not been determined from the expansion method.

The effective surface tension description was originally developed for the 
cosmological quark-hadron transition, where a negative curvature term is 
also present. In the present paper this
approach was generalized to the zero temperature, high 
baryon and electron chemical potential limit.
For both models the conclusions are qualitatively the same. It is the form 
and location, that relative to the size and spacing, become mostly affected
by inclusion of additional finite size terms.
The reason is, 
as shown in Fig.\ \ref{fig4}, that the minimized sum of 
finite size and Coulomb energies differ very little between the different 
geometries, especially between rod and drop like structures. Therefore even 
small changes in the finite size terms may significantly displace the location
and even presence of a given geometry.

This large sensitivity make calculations involving a detailed understanding 
of the crystalline mixed phase unreliable at present, even for the three 
simple geometries. But it also tells us that the wide range of glitch 
phenomena observed in pulsars may be closely related to restructuring in the
solid crystalline region of the pulsar.   

\acknowledgments{M.B.C. wishes to thank Lawrence Berkeley National Laboratory
for hospitality during a three months visit to the Nuclear Theory Group
in the fall of 1996. This work was supported in part by the
Director, Office of Energy Research,
Office of High Energy
and Nuclear Physics,
Division of Nuclear Physics,
of the U.S. Department of Energy under Contract
DE-AC03-76SF00098.}

\begin{appendix}
\section{THE EFFECTIVE SURFACE TENSION}

In contrast to the cosmological transition studied in Ref.\ \cite{potvin93} 
the pressures of the two phases and the surface tension are now functions of
two variables, namely the baryon chemical potential $\mu_n$, and the electron 
chemical potential $\mu_e$.

By $\mu^c_n$, $\mu^c_e$ and $P_c$ we denote the values of the chemical 
potentials and pressure
when the two phases are in equilibrium.
In the proximity of the transition, the hadron and quark pressures can be 
written as
\begin{equation}
 P_h(\mu_n,\mu_e)=P_c-n_{n,h}(\mu^c_n-\mu_n)
                  -n_{e,h}(\mu^c_e-\mu_e)
\end{equation}
and
\begin{equation}
 P_q(\mu_n,\mu_e)=P_c-n_{n,q}(\mu^c_n-\mu_n)
                  -n_{e,q}(\mu^c_e-\mu_e),
\end{equation}
where $n_{n,i}=\left ( \frac{\partial P_i}{\partial \mu_n}  
               \right )_{\mu^c_n,\mu^c_e}$ is the baryon number density
in each of the two phases, and $n_{e,i}=\left ( \frac{\partial P_i}{\partial 
 \mu_e} \right )_{\mu^c_n,\mu^c_e}$ the corresponding electron number 
density.

The pressure difference becomes
\begin{equation}
\label{dp}
 \Delta P=(n_{n,q}-n_{n,h})(\mu^c_n-\mu_n)+(n_{e,q}-n_{e,h})
 (\mu^c_e-\mu_e),
\end{equation}
where the last term vanishes because of the homogeneous electron distribution.

The surface tension is expanded as 
\begin{equation}
\label{sigmatay}
 \sigma(\mu_n,\mu_e)\simeq \sigma(\mu^c_n,\mu^c_e)-\sigma'_n(\mu^c_n-\mu_n)
                     -\sigma'_e(\mu^c_e-\mu_e),
\end{equation}
where $\sigma'_i$ is the partial derivative with respect to $\mu_i$.
Since $\mu_e \ll \mu_n$ we may ignore the third term compared to the second
term.

Combining Eqs.\ (\ref{dp}) and (\ref{sigmatay}) with the Laplace condition for
mechanical (meta)stability,
\begin{equation}
 P_h-P_q=\frac{(d-1)\sigma(\mu_n,\mu_e)}{r},
\end{equation}
generalized to also include cylindrical and planar geometries give a form for
the surface tension, Eqs.\ (\ref{sigma_r}) and (\ref{delta}), similar to 
the one in Ref.\ \cite{potvin93}.

\end{appendix}


\begin{references}
\bibitem{remark1} When we allow the zero-point energy to change sign from one
bulk phase to the other, we obviously break the symmetry between the quark
and hadronic slabs at $\chi =0.5$. But one should keep in mind that the 
approximation altogether, of treating one phase as a bulk phase containing a 
geometry of the other phase, becomes poor when the volume fractions approaches 
0.5.
\end{references}
\end{document}